\documentclass[floatfix,aps,prd,nofootinbib,amsmath,amssymb,amsfonts]{revtex4}

\usepackage{graphicx}

\usepackage{bm}

\newcommand{\be}{\begin{equation}}
\newcommand{\ee}{\end{equation}}
\newcommand{\ba}{\begin{eqnarray}}
\newcommand{\ea}{\end{eqnarray}}

\newcommand{\del}{\nabla}

\newcommand{\p}{\partial}

\newcommand{\D}{\tilde{\nabla}}

\newcommand{\ep}{\varepsilon}
\renewcommand{\a}{\mu}

\begin{document}

\title{Is the Universe homogeneous?}

\author{Roy Maartens}

\affiliation{Department of Physics, University of Western Cape, Cape Town 7535, South Africa\\ Institute of Cosmology \& Gravitation, University of Portsmouth, Portsmouth~PO1~3FX, UK}

\begin{abstract}

The standard model of cosmology is based on the existence of homogeneous surfaces as the background arena for structure formation. Homogeneity underpins both general relativistic and modified gravity models and is central to the way in which we interpret observations of the CMB and the galaxy distribution. However, homogeneity cannot be directly observed in the galaxy distribution or CMB, even with perfect observations, since we observe on the past lightcone and not on spatial surfaces. We can directly observe and test for isotropy, but to link this to homogeneity, we need to assume the Copernican Principle. First, we discuss the link between isotropic observations on the past lightcone and isotropic spacetime geometry: what observations do we need to be isotropic in order to deduce spacetime isotropy? Second, we discuss what we can say with the Copernican assumption.
The most powerful result is based on the CMB: the vanishing of the dipole, quadrupole and octupole of the CMB is sufficient to impose homogeneity. Real observations lead to near-isotropy on large scales -- does this lead to near-homogeneity? There are important partial results, and we discuss why this remains a difficult open question. Thus we are currently unable to prove homogeneity of the Universe on large-scales, even with the Copernican Principle. However we can use observations of the CMB, galaxies and clusters to test homogeneity itself.

\end{abstract}

\maketitle

\section{INTRODUCTION}

The standard model of the Universe -- the LCDM ``concordance" model -- is homogeneous, with structure formation described via perturbations. Given the assumption of homogeneity, and if GR correctly describes gravity, then the acceleration of the Universe is driven by dark energy. The homogeneous LCDM model is highly successful -- a simple, predictive model that is compatible with all observations up to now. However, there is still no satisfactory description of the dark energy that is central to this model. This motivates the need to probe the foundations of the model. We can probe the assumption that GR holds on cosmological scales, by investigating modified gravity theories and by devising consistency tests of GR. This probe is only effective if we assume homogeneity. Alternatively, we can assume that GR holds and probe the assumption of homogeneity (see also \cite{ClaMaa10,Ellis:2011hk}).

\begin{center}
\begin{figure}[htb!]
\vspace*{-2.5cm}
\includegraphics[width=\columnwidth]{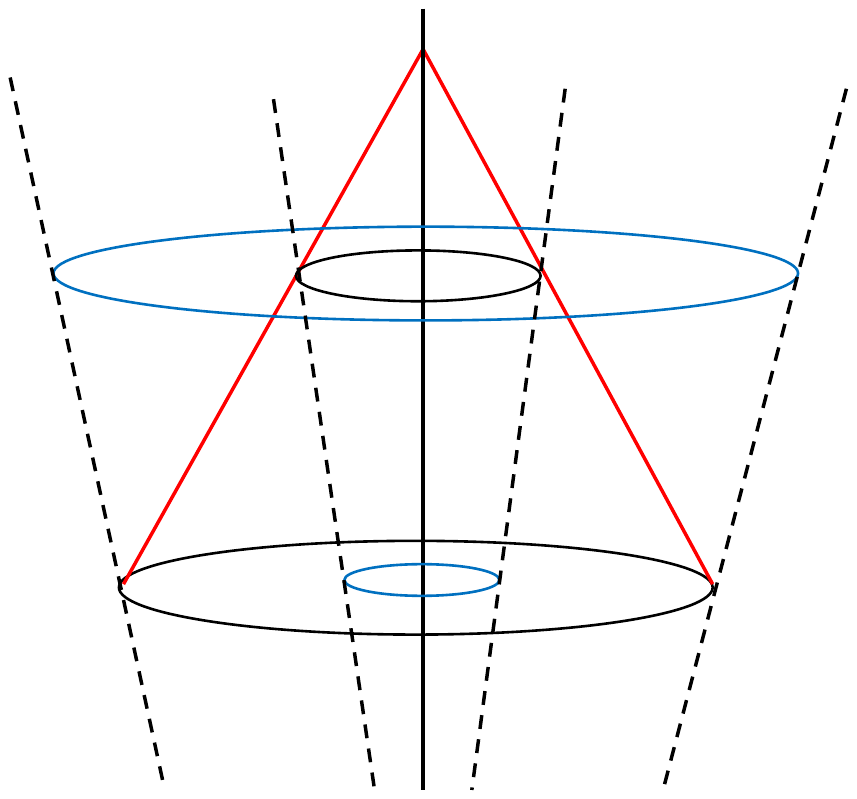}\vspace*{-15.5cm}
\caption{We observe down the past lightcone and therefore cannot directly confirm homogeneity.}\label{fig1}
\end{figure}
\end{center}

A common misconception is that ``homogeneity is obvious from the CMB and the galaxy distribution". In fact, {\em we cannot directly observe or test homogeneity} -- since we observe down the past lightcone, and not on spatial surfaces that intersect that lightcone (see Fig. \ref{fig1}). We only see the CMB on a 2-sphere at one redshift, and galaxy surveys give us the galaxy distribution on 2-spheres of constant redshift. There are interesting and important analyses of the observed galaxy distribution to probe statistical homogeneity (e.g. \cite{Sarkar09,Sylos11}), but these effectively assume an FLRW background geometry.

What we can directly test for is {\em isotropy of observations}. This then raises an important, but often overlooked question:  {\em what observational quantities need to be isotropic in order to enforce isotropy of the geometry?} This question is addressed in Sec. \ref{isosec}. To answer this question we need a fully nonlinear analysis, since we cannot assume a priori any symmetric background spacetime. For observations of the galaxy distribution, the answer is -- we need isotropic angular distances, number counts, bulk velocities and lensing. Isotropy of the CMB by contrast does {\em not} in itself enforce spacetime isotropy.

In order to link isotropy to homogeneity, we have to assume the Copernican Principle (CP), i.e. that we are not at a special position in the Universe. The CP is not observationally based; it is an expression of the intrinsic limitation of observations from one spacetime location. We consider in Sec. \ref{homosec} what can be done with the CP. If we have isotropy along one worldline, based on the observed galaxy distribution, then the CP leads to homogeneity. A more powerful result is that {\em homogeneity follows if all observers see isotropic angular distances up to third order in redshift.} The strongest basis for homogeneity comes from the CMB. This is often considered to be obvious -- but it is far from straightforward to show that homogeneity of the metric follows if all observers see isotropic CMB. The proof requires the general nonlinear Einstein-Liouville equations. Remarkably, it is not necessary to assume full CMB isotropy for each observer -- it is enough that each observer {\em sees isotropy in the CMB only up to the octupole.} This is the most powerful observational basis for homogeneity currently known.

Of course the real CMB is not exactly isotropic, but nearly isotropic. Does it follow from near-isotropy that the Universe is nearly homogeneous, i.e. perturbed FLRW? This has so far only been shown with further assumptions on the gradients and time derivatives of CMB multipoles.

It is important to stress from the outset that there are two fundamental limitations:

\begin{itemize}\itemsep=-4pt
\item
Isotropy and homogeneity of observables can only be meaningfully defined on large enough scales -- and the nature of the transition scale is only poorly understood.
\item
Isotropy and homogeneity of observables can only be meaningfully defined in an average sense -- and the problem of how to average in general relativity (and other metric theories that are intrinsically nonlinear) remains unsolved.
\end{itemize}
These unresolved issues are of crucial importance in cosmology, but they are not discussed here; instead, observations are treated as idealized.


\section{WHAT IS THE BASIS FOR ISOTROPY?}\label{isosec}

Here we consider the situation when the CP is {\em not} adopted. What is the observational basis for showing that spacetime is isotropic about the worldline of a single observer? Isotropy is directly observable and the best example is the CMB, which is isotropic about us to $\sim 10^{-5}$ (after the dipole is interpreted as due to our motion relative to the cosmic frame, and removed by a boost). Observations of the galaxy distribution do not have the same precision, but there is no evidence for anisotropy.
First we look at observations of matter and then of the CMB.

\subsection{Isotropic matter distribution on the past lightcone of one observer}

The dominant cosmological components -- cold dark matter
and dark energy -- have not been independently observed. Unlike baryonic matter, the dark components are up to now only manifest via their
gravitational effect. The distribution of dark matter is mapped by weak lensing
surveys. But to relate the measured projected
potential on the sky at each redshift to the dark matter, we
require a specific model, such as a perturbed FLRW model. The dark
matter 4-velocity is usually assumed to be aligned with that of
baryonic matter -- but this is also based on a perturbed FLRW model.

This unavoidably means that we must impose a
model for these dark components -- not merely their physical
properties, but how they relate spatially to observed matter -- in
order to determine their distribution via cosmological observations.
A starting point \cite{ClaMaa10} is to assume that the CDM 4-velocity is the same
as the baryonic 4-velocity, and that we know the primordial ratio of
CDM density to baryonic density, as well as the bias factor
that relates the concentrations of CDM and baryons in
clustered matter
\begin{equation}\label{cdm from b}
\rho_{\rm c}~\mbox{known from}~\rho_{{\rm b}} ~~\mbox{and}~~ u_{{\rm b}}^a =
u_{\rm c}^a :=u^a\,.
\end{equation}
If there is a modified gravity theory that avoids the need for CDM, then we do not need (\ref{cdm from b}) -- but we are likely to need other assumptions on the extra degrees of freedom that mimic CDM.

If dark energy is
in the form of $\Lambda$, then we need to assume
that its value is known from non-cosmological
physics:
 \be \label{lam given}
\Lambda~\mbox{known independently of cosmological observations.}
 \ee
For quintessence and other more complicated forms of dark energy, we would need to assume how the dark energy field is distributed in spacetime -- since we are not assuming a FLRW geometry a priori. If there is a modified gravity theory that avoids the need for dark energy, then (\ref{lam given}) is not needed, but assumptions will likely be necessary on the extra degrees of freedom that mimic dark energy.

Given the assumptions on the dark components,
what can we say about spacetime geometry if the matter distribution is isotropic on the past lightcone of the observer? Which observables need to be isotropic in order to deduce isotropic geometry? It turns out that 4 independent observables on the lightcone are exactly enough to impose isotropy of spacetime.
The original result for a baryonic universe \cite{7,7b,7a} may be updated to include CDM and $\Lambda$ \cite{ClaMaa10}, incorporated via (\ref{cdm from b}) and (\ref{lam given}):
\begin{quote}
\textbf{\em Isotropy of matter distribution on the lightcone $\rightarrow$ isotropy of spacetime geometry}

If one fundamental observer comoving with the matter measures isotropy of (a)~angular diameter distances, (b)~number counts, (c)~bulk  velocities, and (d)~lensing, in an expanding dust universe with $\Lambda$, then the spacetime is isotropic about the observer's worldline.
\end{quote}

Note that isotropy of bulk velocities is equivalent to vanishing transverse velocities (proper motions) on the observer's sky. Isotropy of lensing means that there is no distortion of images, only magnification.

\begin{center}
\begin{figure}[htb!]
\vspace*{-2.5cm}
\includegraphics[width=\columnwidth]{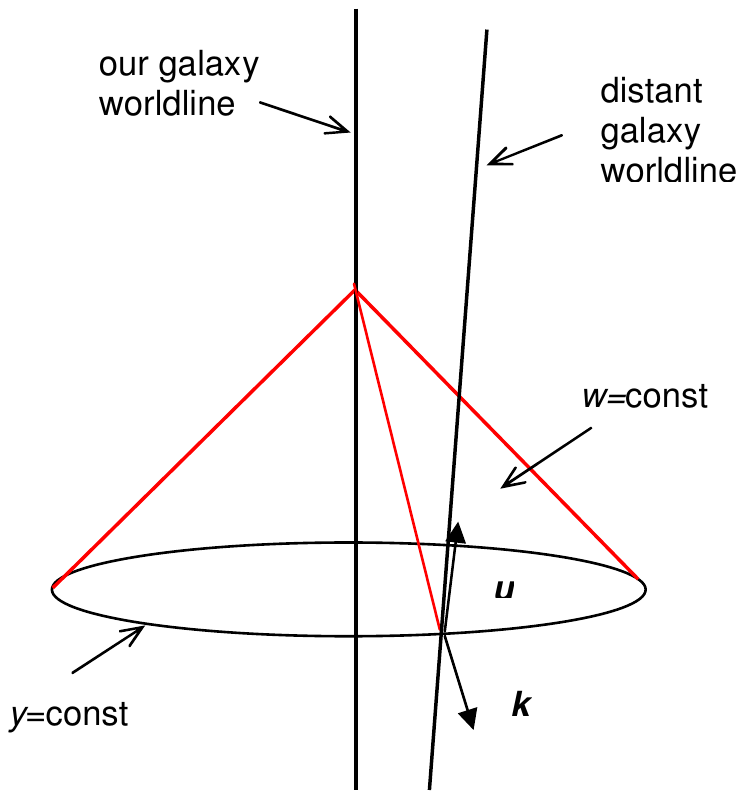}\vspace*{-15.5cm}
\caption{Observational coordinates based on the past lightcones of the observer's worldline.}\label{fig2}
\end{figure}
\end{center}

The proof of this result requires a non-perturbative approach -- there is no background to perturb around. Since the data is given on the past lightcone of the observer,  we need the full general metric, adapted to the past lightcones of the observer worldline ${\cal C}$. We define observational coordinates $x^\a=(w,y,\theta,\phi)$, where $x^P=(\theta,\phi)$ are the celestial coordinates, $w=\,$const are the past light cones on ${\cal C}$ ($y=0$), normalized so that $w$ measures proper time along ${\cal C}$, and $y$ measures distance down the light rays $(w,\theta,\phi)=\,$const (see Fig. \ref{fig2}). A convenient choice for $y$ is $y=z$ (redshift) on the lightcone of here-and-now, $w=w_0$, and then keep $y$ comoving with matter off the initial lightcone, so that $u^y=0$. Then the matter 4-velocity and the photon wave-vector are
\begin{equation}\label{ukobs}
u^\a=(1+z) (1,0, V^P)\,,~~ k_\a=w_{,\a}\,, ~~ 1+z=u_\a k^\a,
\end{equation}
where $V^P=\mathrm{d} x^P/\mathrm{d} w$ are the transverse velocity components on the observer's sky.

The metric of an arbitrary spacetime in observational coordinates is
\begin{eqnarray}
\mathrm{d} s^2 &=&\! -A^2\mathrm{d} w^2+ 2B\mathrm{d} w \mathrm{d} y+ 2C_P\mathrm{d} x^P \mathrm{d} w+D^2(\!\mathrm{d}\Omega^2+ L_{PQ}\mathrm{d} x^P \mathrm{d} x^Q )~~ \label{} \\
A^2 &=& (1+z)^{-2}+2C_PV^P+ g_{PQ}V^P V^Q\,,~~~ B={\mathrm{d} v \over \mathrm{d} y},
\end{eqnarray}
where $v$ is a null affine parameter, $D$ is the angular diameter distance, and $L_{PQ}$ determines the lensing distortion of images via the shear of lightrays,
\begin{equation}\label{shear}
\hat\sigma_{PQ}= {D^2 \over 2B} {\p L_{PQ} \over \p y}.
\end{equation}
The number of galaxies in a solid angle $\mathrm{d}\Omega$ and a null distance increment $\mathrm{d} v=B{\rm d}y$ is
\begin{equation}\label{nc}
\mathrm{d} N = Sn(1+z)D^2B \mathrm{d}\Omega \mathrm{d} y\,,
\end{equation}
where $S$ is the selection function and $n=\rho_{\rm m}/m$ is the number density.

Before specializing to isotropic observations, we identify how the observations in general and in principle determine the geometry of the past light cone $w=w_0$ of here-and-now, where $y=z$:

\begin{itemize}\itemsep=-4pt
\item
Given the intrinsic properties and evolution of sources,
observations in principle determine:
(a)~the angular diameter distance $D(w_0,z,x^P)$,
(b)~the lensing
distortion of images, $L_{PQ}(w_0,z,x^R)$. Thus the metric components $g_{PQ}$ on $w=w_0$ are determined.
\item
Given (\ref{cdm from b}) and the selection function, the number counts $N(w_0,z,x^P)$ in principle determine $B(w_0,z,x^P)\rho_{\rm m}(w_0,z,x^P)$.
\item
In principle, observations over extended timescales determine the
instantaneous transverse velocities $V^P(w_0,z,x^Q)$ of discrete
sources. Transverse velocities of clusters are in principle determined by the polarization of scattered CMB photons \cite{Sunyaev:1980nv}.
\end{itemize}

It follows that, in principle and for idealized observations:
\begin{equation}\label{id}
\mbox{Idealized data}~ \Rightarrow ~ \{u^\mu\,,B\rho_{\rm
  m}\,,g_{PQ}\}~\mbox{on}~ w=w_0.
\end{equation}
But this is insufficient to determine the geometry of the
past lightcone, because we need $g_{0P}=C_P$ and we cannot separate out $B$ and
$\rho_{\rm m}$. Without
gravitational field equations, we are unable to fully
determine the spacetime and matter on the past lightcone, even
assuming perfect information from discrete-source observations. As a consequence, it is also impossible to test
gravity theories directly:
\begin{quote}\label{th:test gr}
\textbf{\em Cosmological testing of gravity theories}\\
Even with perfect
observations, we cannot
determine the spacetime geometry and matter on our past
lightcone without gravitational field equations. Thus
observations cannot directly test GR on
cosmological scales, or test any alternative theories of
large-scale gravity, without making assumptions about the spacetime geometry.
\end{quote}

If GR holds, then (\ref{id}) is exactly what is needed \cite{7,7b}:
\begin{quote}
\textbf{\em Matter observations $\rightarrow$ metric and matter on lightcone}

Observational data (\ref{id}) is exactly the information needed for Einstein's equations to determine $B$ and $C_P$ on $w=w_0$, so that the metric and matter are fully determined on the lightcone. Then the past Cauchy development of this data determines $g_{\a\nu},u^\a,\rho_{\rm m}$ in the interior of the past lightcone.
\end{quote}

If the matter observations are isotropic, then we can prove isotropy of spacetime and matter \cite{7,7b}:
\begin{quote}
\textbf{\em Isotropy of lightcone matter distribution for one observer $\rightarrow$ isotropic spacetime}

In an expanding dust region with $\Lambda$, if one fundamental observer measures isotropic angular diameter distances, number counts, bulk velocities, and lensing,
\begin{equation}\label{misob}
{\p D \over \p x^P}= {\p N\over \p x^P}=V^P=L_{PQ}=0\,,
\end{equation}
then the spacetime and matter are isotropic, i.e. the region is Lemaitre-Tolman-Bondi (LTB).
\end{quote}

It is an open question how this result translates to the case of almost-isotropy of observations, i.e. {\em does this lead to almost-isotropy of the spacetime?}

\subsection{Isotropic CMB for one observer}

If the CMB is isotropic for one observer comoving with the matter, then along the observer's worldline ${\cal C}$ we have [see (\ref{iso})]
\begin{equation}
f(x_{\cal C},p)=F(x_{\cal C},E),~~F_{\a_1 \cdots \a_\ell}\big|_{\cal C}=0 =\dot{F}_{\a_1 \cdots \a_\ell}\big|_{\cal C}~\mbox{for}~ \ell \geq 1\,.
\end{equation}
In other words, all covariant multipoles of the distribution function beyond the monopole [see (\ref{r3})], and their time derivatives, must vanish along ${\cal C}$. By (\ref{em3}), the radiation momentum density (from the dipole) and anisotropic stress (from the quadrupole)  vanish:
$q_{\rm r}^\a|_{\cal C}=0=\pi_{\rm r}^{\a\nu}|_{\cal C}$. However, without the Copernican assumption, we are not able to deduce directly the vanishing of spatial derivatives $\D_\nu F_{\a_1 \cdots \a_\ell}\big|_{\cal C}$, and then we cannot show isotropy of the spacetime geometry about ${\cal C}$.

Isotropy of the CMB alone is not sufficient, since the matter could in principle be anisotropic, even if this is physically unnatural. In order to rule out artificial counter-examples, it would be sufficient to {\em characterize the minimal matter isotropy that combines with CMB isotropy to give geometric isotropy.}

If we adopt the Copernican Principle, then the CMB alone leads to a powerful result, as discussed in the next section.

\section{WHAT IS THE BASIS FOR SPATIAL HOMOGENEITY? }\label{homosec}

Homogeneity cannot be directly observed -- we are effectively unable to move away in cosmic time or distance from here-and-now and hence cannot probe spatial variations on constant time slices; effectively, our observations only access the past lightcone of here-and-now. Direct observation thus cannot distinguish between an evolving homogeneous distribution of matter and inhomogeneity with a different time evolution -- since the past light cone only accesses a 2-sphere in each constant-time slice (see Fig. \ref{fig1}). Thus we are forced to adopt the Copernican assumption. We first consider matter observations, then an exactly isotropic CMB, and finally the case of an almost-isotropic CMB.

\subsection{Isotropic matter observations for all observers}

If all observers see isotropy of the 4 matter observables, then we have geometric isotropy along all worldlines and thus homogeneity follows.

\begin{quote}
\textbf{\em Isotropy of lightcone matter distribution for all observers $\rightarrow$ FLRW}

In an expanding dust region with $\Lambda$, if all fundamental observers measure isotropic angular diameter distances, number counts, bulk velocities, and lensing, then the spacetime is FLRW.
\end{quote}

This is an improved form of the Cosmological Principle -- based on isotropy of specific observables and not on assumed geometric isotropy.

In fact there is a much stronger statement than this, based only on one observable, the angular diameter distance, and only for small redshifts \cite{HP}:

\begin{quote}
\textbf{\em Isotropic distances up to $O(z^3)$ for all observers $\rightarrow$ FLRW}

If all fundamental observers in an expanding spacetime region measure isotropic angular diameter distances up to third-order in a redshift series expansion, then the spacetime is FLRW in that region.
\end{quote}

A covariant proof of this \cite{Cthesis,ClaMaa10} is based on a series expansion in a general spacetime, using the method of Kristian and Sachs \cite{KS}. The redshift may be expanded in terms of the angular diameter distance:
 \ba
z&=&\big[k^\a k^\nu \nabla_\a u_\nu\big]_OD+ {1\over2} \big[k^\a k^\nu k^\alpha \nabla_\a \nabla_\nu u_\alpha\big]_OD^2 \nonumber\\ && +{1\over6}\Big[k^\a k^\nu k^\alpha k^\beta \nabla_\a \nabla_\nu \nabla_\alpha u_\beta +{1\over2} k^\a k^\nu k^\alpha k^\beta R_{\alpha\beta} \nabla_\a u_\nu\Big]_OD^3 +O(D^4).
 \ea
Here all terms are evaluated at the observer $O$ in the unit direction $e^\a$ of the lightray, where [see (\ref{ukobs})]:
 \be
k^\a=-(1+z)(u^\a +e^\a), ~~~u_\a e^\a=0,~ e_\a e^\a=1.
 \ee
The covariant derivative of the 4-velocity is decomposed as in (\ref{kin}).
The $O(D)$ coefficient is the observed Hubble rate (generalizing the FLRW quantity),
 \be
H^\mathrm{obs}_{0}=\big[k^\a k^\nu\del_\a u_\nu\big]_0=\Big[{1\over3}\Theta +A_\a e^\a+ \sigma_{\mu\nu}e^\a e^\nu \Big]_O.
 \ee
Isotropy at lowest order, for all observers $O$, thus enforces vanishing acceleration and shear
 \be \label{accsh}
A_\a = 0 =\sigma_{\a \nu}.
 \ee
Using (\ref{accsh}), the next coefficient becomes
 \ba
\big[k^\a k^\nu k^\alpha \nabla_\a \nabla_\nu u_\alpha\big]_O ={1\over6} \Big[ \big(2\Theta^2+\rho_{\rm m}-2\Lambda\big)-2\D_\a\Theta e^\a+ \big(E_{\a\nu}+ \omega_{\langle\a} \omega_{\nu\rangle}\big)e^\a e^\nu \Big]_O,
 \ea
where $\D_\a$ is the covariant spatial derivative (\ref{dd}), $E_{\a\nu}$ is the electric part of the Weyl tensor (\ref{gem}), and the angled brackets denote the spatial tracefree part (\ref{pstf}). Isotropy at $O(D^2)$ therefore imposes homogeneity of the expansion rate,
$ \D_\a \Theta =0$,
and the condition $ E_{\a\nu}=- \omega_{\langle\a} \omega_{\nu\rangle}$. However, this condition is identically satisfied by virtue of (\ref{accsh}) and the shear propagation equation (\ref{e5}). The constraint equations (\ref{c2}), (\ref{c3}) then show that
$\mbox{curl}\,\omega_\a=0$ and $H_{\a\nu}=\D_{\langle\a}\omega_{\nu\rangle}$,
where $H_{\a \nu}$ is the magnetic Weyl tensor. The $O(D^3)$ coefficient imposes further constraints, and we find that $0=\D_\a\rho_{\rm m}=\omega_\a=E_{\a\nu}=H_{\a\nu}$. Putting everything together, we have a covariant characterization of FLRW.

\subsection{Exactly isotropic CMB for all observers}

It is commonly assumed that isotropy of the CMB for all observers leads obviously to FLRW, without the need for any proof. In fact, the proof is far from obvious, and requires a nonlinear analysis of the general Einstein-Liouville equations.
The starting point is a pioneering mathematical result by Ehlers, Geren and Sachs \cite{EGS}. They assumed that the only source of the gravitational field was the radiation, i.e. they neglected matter and $\Lambda$. This can be generalized to include self-gravitating matter and dark energy \cite{ClaMaa10} (extending \cite{2,TreEll71,FerMorPor99,CB,CC,Rasanen:2009mg}):

\begin{quote}
\textbf{\em CMB isotropy for all observers $\rightarrow$ FLRW}\\
In a region, if
\vspace*{-4pt}
\begin{itemize}\itemsep=-4pt
\item collisionless radiation is exactly isotropic,
\item the radiation four-velocity is geodesic and expanding,
\item there is dust matter and dark energy in the form of $\Lambda$, quintessence or a perfect fluid,
\end{itemize}
then the metric is FLRW in that region.
\end{quote}

The fundamental 4-velocity $u^\a$ is the radiation 4-velocity, which has zero 4-acceleration and positive expansion:
\begin{equation}\label{av}
A_\a=0\,,~~\Theta >0\,.
\end{equation}
Isotropy of the radiation distribution about $u^\a$ means that photon peculiar velocities are isotropic for comoving observers; thus in momentum space, the photon distribution depends only on components of the 4-momentum $p^\a$ along $u^\a$, i.e., on the photon energy $E=-u_\a p^\a$:
\begin{equation}\label{iso}
f(x,p)=F(x,E),~~F_{\a_1 \cdots \a_\ell}=0~\mbox{for}~ \ell \geq 1\,.
\end{equation}
In other words, all covariant multipoles of the distribution function beyond the monopole, defined in (\ref{r3}), must vanish. In particular, (\ref{em3}) shows that:
\begin{equation}
\label{qpv}
q_{\rm r}^\a=0=\pi_{\rm r}^{\a\nu}\,.
\end{equation}
Equation (\ref{iso}) also implies that the radiation brightness octupole $\Pi_{\a\nu\alpha}$ and hexadecapole $\Pi_{\a\nu\alpha\beta}$ are zero. These are source terms in the anisotropic stress evolution equation, which is the $\ell=2$ case of (\ref{r26}). The general nonlinear form of the $\pi_{\rm r}^{\a\nu}$ evolution equation is \cite{2,Maartens:1998xg}
\begin{eqnarray}
&&\dot{\pi}_{\rm r}^{\langle \a\nu \rangle}+{{4\over3}}\Theta
\pi_{\rm r}^{\a\nu } +{{8\over15}}\rho_{\rm r}\sigma^{\a\nu }+
{{2\over5}}\D^{\langle \a}q_{\rm r}^{\nu\rangle} +2 A^{\langle \a} q_{\rm r}^{\nu\rangle} -2\omega^{\alpha}\ep_{\alpha \beta }{}{}^{(\a}
\pi_{\rm r}^{\nu) \beta} ~~~~~
\nonumber\\
&& ~~~~~~~~  +{{2\over7}}\sigma_{\alpha}{}^{\langle
\a}\pi_{\rm r}^{\nu\rangle \alpha} +{8\pi\over35}\D_{\alpha}
\Pi^{\a\nu \alpha}  -{32\pi\over315} \sigma_{\alpha \beta } \Pi^{\a\nu \alpha \beta }
= 0. \label{nl8}
\end{eqnarray}
Isotropy removes all terms on the left except the third, and thus enforces a shear-free expansion of the fundamental congruence:
\begin{equation}\label{sv}
\sigma_{\a\nu}=0\,.
\end{equation}

We can also show that $u^\a$ is irrotational as follows. Together with (\ref{av}), momentum conservation for radiation, i.e., (\ref{e3i}) with $I=r$, reduces to
\begin{equation}\label{mv}
\D_\a \rho_{\rm r}=0\,.
\end{equation}
Thus the radiation density is homogeneous relative to fundamental observers.
Now we invoke the exact nonlinear identity for the covariant curl of the gradient, (\ref{ri1}):
\begin{equation}\label{}
\mbox{curl}\, \D_\a \rho_{\rm r} = - 2\dot \rho_{\rm r} \omega_\a~ \Rightarrow~ \Theta \rho_{\rm r} \omega_\a =0\,,
\end{equation}
where we have used the energy conservation equation (\ref{e1i}) for radiation. By assumption $\Theta >0$, and hence we deduce that the vorticity must vanish:
\begin{equation}\label{vv}
\omega_\mu =0\,.
\end{equation}
Then we see from the curl shear constraint equation (\ref{c3}) that the magnetic Weyl tensor must vanish:
\begin{equation}\label{hv}
H_{\a\nu}=0\,.
\end{equation}

Furthermore, (\ref{mv}) actually tells us that the expansion must also be homogeneous. From the radiation energy conservation equation (\ref{e1i}), and using (\ref{qpv}), we have $\Theta=-3{\dot\rho_{\rm r}}/{4\rho_{\rm r}}$.
On taking a covariant spatial gradient and using the commutation relation (\ref{timespace}), we find
\be \label{tv}
\D_\a\Theta=0\,.
\ee
Then the shear divergence constraint, (\ref{c2}), enforces the vanishing of the total momentum density in the fundamental frame,
\begin{equation}\label{qv}
q^\a := \sum_I q_I^\a =0~ \Rightarrow ~\sum_I \gamma_I^2(\rho^*_I + p^*_I)v_I^\a=0  \,.
\end{equation}
The second equality follows from (\ref{t6}), using the fact that the baryons, CDM and dark energy (in the form of quintessence or a perfect fluid) have vanishing momentum density and anisotropic stress in their own frames, i.e.,
\begin{equation}\label{perf}
q_I^{*\a}=0= \pi_I^{*\a\nu}\,,
\end{equation}
where the asterisk denotes the intrinsic quantity (see Appendix A). If we include other species, such as neutrinos, then the same assumption applies to them. Except in artificial situations, it follows from (\ref{qv}) that
\begin{equation}\label{pvan}
v_I^\a=0\,,
\end{equation}
i.e., the bulk peculiar velocities of matter and dark energy [and any other self-gravitating species satisfying (\ref{perf})] are forced to vanish -- all species must be comoving with the radiation.

The comoving condition (\ref{pvan}) then imposes the vanishing of the total anisotropic stress in the fundamental frame:
\begin{equation}\label{piv}
\pi^{\a\nu}:= \sum_I \pi_I^{\a\nu} =\sum_I \gamma_I^2(\rho^*_I + p^*_I)v_I^{\langle \a}v_I^{\nu \rangle}=0  \,,
\end{equation}
where we used (\ref{t7}), (\ref{perf}) and (\ref{pvan}). Then the shear evolution equation (\ref{e5}) leads to a vanishing electric Weyl tensor
\begin{equation}\label{evan}
E_{\a\nu}=0\,.
\end{equation}
Equations (\ref{qv}) and (\ref{piv}), now lead via the total momentum conservation equation (\ref{e3}) and the $E$-divergence constraint (\ref{c4}), to homogeneous total density and pressure:
\begin{equation}\label{rpv}
\D_\a\rho = 0 = \D_\a p\,.
\end{equation}

Equations (\ref{av}), (\ref{sv}), (\ref{hv}), (\ref{tv}), (\ref{qv}), (\ref{piv}) and (\ref{rpv}) constitute a covariant characterization of an FLRW spacetime. This establishes the EGS result, generalized from the original to include self-gravitating matter and dark energy, and presented in a fully covariant form. It is straightforward to include other species such as neutrinos. The critical assumption needed for all species is the vanishing of the intrinsic momentum density and anisotropic stress, i.e., (\ref{perf}). Equivalently, the energy-momentum tensor for the $I$-component should have perfect fluid form in the $I$-frame. The isotropy of the radiation and the geodesic nature of its 4-velocity then enforce the vanishing of (bulk) peculiar velocities $v_I^\a$. We emphasize that one does {\em not} need to assume that the matter or other species are comoving with the radiation -- it follows from the assumptions on the radiation.

In fact this result can be dramatically strengthened on the basis of a theorem by Ellis, Treciokas and Matravers \cite{Ellis:1983}: we do not need vanishing of all multipoles, but only the dipole, quadrupole and octupole! The key step is to show that the shear vanishes, without having zero hexadecapole -- the quadrupole evolution equation (\ref{nl8}) no longer automatically gives $\sigma_{\a\nu}=0$, and we need to find another way to show this. The ETM trick is to return to the Liouville multipole equation (\ref{r25}). The $\ell=2$ multipole of this equation, with $F_\a=F_{\a\nu}= F_{\a\nu\alpha}=0$, gives
\begin{equation}\label{}
{12 \over 63}{\p \over \p E}\left( E^5\sigma^{\a\nu}F_{\a\nu\alpha\beta} \right)+ E^5 {\p F\over \p E}\,\sigma_{\alpha\beta}=0\,.
\end{equation}
We integrate over $E$ from 0 to $\infty$, and use the convergence property $E^5 F_{\a\nu\alpha\beta} \to 0$ as $E \to \infty$. This gives
\begin{equation}\label{}
\sigma_{\alpha\beta} \int_0^\infty E^5 {\p F\over \p E}\mathrm{d} E=0\,.
\end{equation}
Integrating by parts, the integral reduces to $-5\int_0^\infty E^4 F \mathrm{d} E$. Since $F>0$, the integral is strictly negative, and thus we arrive at vanishing shear, $\sigma_{\a\nu}=0$. Then the proof above proceeds as before, incorporating matter and dark energy to extend the ETM result \cite{ClaMaa10}:

\begin{quote}
\textbf{\em CMB partial isotropy for all observers $\rightarrow$ FLRW}

In a region, if
\vspace*{-4pt}
\begin{itemize}\itemsep=-4pt
\item collisionless radiation has vanishing dipole, quadrupole and octupole,
\begin{equation}\label{}
F_\a= F_{\a\nu}= F_{\a\nu\alpha}=0\,,
\end{equation}
\item the radiation four-velocity is geodesic and expanding,
\item there is dust matter and dark energy in the form of $\Lambda$, quintessence or a perfect fluid,
\end{itemize}
then the metric is FLRW in that region.
\end{quote}

This is the most powerful observational basis that we have for background homogeneity and thus an FLRW background model.

\subsection{The real universe: almost-isotropic CMB}

In practice we can only observe approximate isotropy (see Fig. \ref{fig3}). Is the previous result stable -- i.e., does almost-isotropy of the CMB lead to an almost-FLRW Universe? This would be the {\em realistic} basis for a spatially homogeneous Universe (assuming the Copernican Principle). It  was shown to be the case, but subject to further assumptions, by \cite{2,Maartens:1994qq}:

\begin{center}
\begin{figure}[htb!]
\includegraphics[width=0.5\columnwidth]{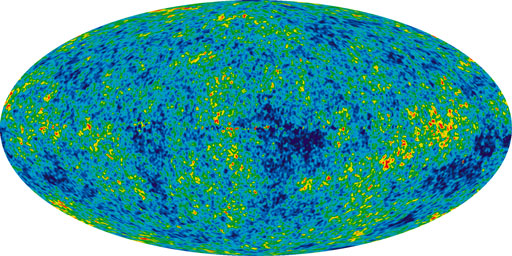}
\caption{The CMB temperature from WMAP 7-year data, showing anisotropies at the level $\sim 10^{-5}$. (Credit: NASA/ WMAP Science Team.)
}\label{fig3}
\end{figure}
\end{center}

\begin{quote}
\textbf{\em A more realistic basis for homogeneity}

In a region of an expanding Universe with dust and cosmological constant, if all observers comoving with the matter measure an almost isotropic distribution of collisionless radiation, and if some of the time and spatial derivatives of the covariant multipoles are also small, then the region is almost FLRW.
\end{quote}

We emphasize that the perturbative assumptions are purely on the photon distribution, not the matter or the metric -- and one has to prove that the matter and metric are then perturbatively close to FLRW. Once again, a nonperturbative analysis is essential, since we are trying to prove an almost-FLRW spacetime, and we cannot assume it a priori.

Almost-isotropy of the photon distribution means that
\begin{equation}\label{}
F_{\a_1 \cdots \a_\ell}(x,E)= {\cal O}(\epsilon), ~~ \ell\geq 1\,,
\end{equation}
where $\epsilon$ is a (dimensionless) smallness parameter.
The brightness multipoles $\Pi_{M_\ell}$ have dimensions of energy density and we therefore normalize them to the monopole $\Pi= \rho_{\rm r}/4\pi$, so that
${\Pi_{M_\ell} / \Pi}= {\cal O}(\epsilon)$.

The task is to show that the relevant kinematical, dynamical and curvature quantities, suitably non-dimensionalized, are ${\cal O}(\epsilon)$. For example, the dimensionful kinematical quantities may be normalized by the expansion, $\sigma_{\a\nu}/\Theta, \omega_\a/\Theta$. The proof then follows the same pattern as the proof above of the exact result -- except that at each stage, we need to show that quantities are ${\cal O}(\epsilon)$ rather than equal zero.

However, in order to show this, we need smallness not just of the multipoles, but also of some of their derivatives. Smallness of the multipoles does not directly imply smallness of their derivatives, and we have to assume this \cite{Rasanen:2009mg,Rasanen:2008be}. It remains a difficult open problem whether these additional assumptions may be removed.
If all observers measure small multipoles, then it may be possible to use almost-isotropy of the matter distribution to show that the time and space derivatives on cosmologically significant scales must also be small.

A number of experiments has been proposed to test the Copernican Principle by looking for violations of isotropy at events down our past lightcone, as discussed in the next section. The almost-isotropic CMB result then gives a framework for probing inhomogeneities via such observations. Indeed, these tests may provide a way of constraining spatial gradients of the low-$\ell$ multipoles.

In addition, it may be possible to strengthen the almost-isotropic result above by proving that it is sufficient for only the first 3 multipoles and their derivatives to be small. This would represent a more realistic foundation for almost-homogeneity.

\section{TESTING HOMOGENEITY}\label{cpsec}

Although we cannot directly probe homogeneity by observations, we can test for violations of homogeneity. If we find no violation, then the indirect evidence for homogeneity is strengthened. However, if even one significant violation is discovered, then homogeneity will have been disproved. There are two broad classes of observational tests of homogeneity, one based on consistency relations that hold in FLRW and the other based on using galaxy clusters as probes of the anisotropy seen in the CMB from positions at cosmological distances down our past lightcone.

\subsection{Consistency of distances and expansion rate}

The effective standard candles provided by supernovae observations lay the basis for a consistency test of homogeneity. There are two geometric effects on distance measurements: the curvature
bends null geodesics and the expansion changes radial
distances. These are coupled in FLRW models via
\begin{equation}
 D_L(z)= \frac{(1+z)}{H_0\sqrt{-\Omega_{K0}}}
\sin{\left(\sqrt{-\Omega_{K0}}\int_0^z {{\rm d} z' \over H(z')/H_0}\right)},
\end{equation}
and then one can combine the Hubble rate and distance data
to find the curvature today:
\begin{equation}\label{omk}
\Omega_{K0} = \frac{H^2(z)[(1+z)D_L'(z)-D_L(z)]^2-(1+z)^4}{H_0^2(1+z)^2D^2_L(z)}.
\end{equation}
This relation is independent of all other cosmological parameters,
including dark energy -- and it is also independent of the theory of gravity. It can be used
at a single redshift to determine $\Omega_{K0}$. Furthermore, it is the basis for a CP test, proposed by
\cite{CBL} -- since $\Omega_{K0}$ is
independent of $z$, we can differentiate (\ref{omk}) to get the consistency
relation:
\begin{eqnarray}
{\cal K}(z) &:=& (1+z)^4 + H^2(z)\left[(1+z)^2\left\{D_L(z)D_L''(z)- D_L^{\prime 2}(z)\right\}+D_L^2(z) \right]
\nonumber\\ &&~
+ (1+z)H(z) H'(z)D_L(z)\left[(1+z) D_L'(z)-D_L(z) \right] \nonumber\\ & =&0 ~~\mbox{for FLRW geometry}. \label{cz test}
\end{eqnarray}
Note that ${\cal K}=0$ for {\em any} FLRW geometry, independent of curvature,
dark energy, matter content, and theory of gravity. In realistic
models we should expect $|{\cal K}(z)| \sim 10^{-5}$, reflecting
perturbations from large-scale structure formation. Significantly larger values indicate a breakdown of homogeneity:
 \be
{\cal K}(z) ~~\mbox{significantly different from 0} ~~\Rightarrow ~~ \mbox{non-FLRW universe.}
 \ee
Carrying out this test should not be more difficult than carrying out
dark energy measurements of $w(z)$ from SNIa data, which require
$H'(z)$ from distance measurements or the second derivative
$D_L''(z)$.

This is the simplest test of homogeneity, and its
implementation should be regarded as a high priority.
Another test involves the time drift of the cosmological redshift \cite{UCE}, but this will only be feasible on a much longer timescale.

\begin{center}
\begin{figure}[htb!]
\vspace*{-2cm}
\includegraphics[width=\columnwidth]{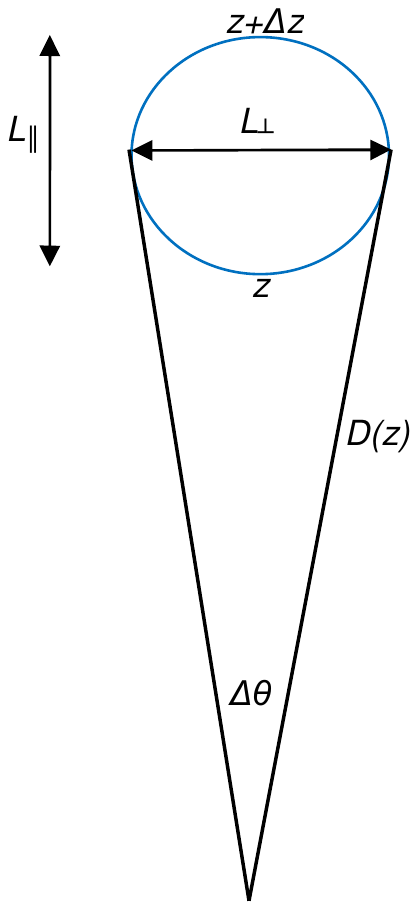}\vspace*{-15.5cm}
\caption{Radial and transverse BAO scales.
}\label{fig5}
\end{figure}
\end{center}

Finally the BAO feature itself provides in principle a homogeneity test. Future large-volume surveys will allow the detection of the BAO scale in both radial and transverse directions. The physical lengths in radial and transverse directions of a feature with redshift extent $\Delta z$ and subtending an angle $\Delta \theta$ are
 \be
L_\|={ \Delta z \over (1+z)H_\parallel (z)},~~~ L_\perp =D(z) \Delta\theta,
 \ee
where $D$ is angular diameter distance and $H_\parallel$ is the expansion rate in the radial direction (see Fig. \ref{fig5}).
The two scales are equal for the BAO feature in an FLRW background, where expansion is isotropic at all points. Any significant disagreement between the radial and transverse BAO scales would signal a breakdown of remote isotropy and thus of homogeneity:
 \be
 {L_\| \over L_\perp}-1 ~~\mbox{significantly different from 0}  ~~\Rightarrow ~~ \mbox{non-FLRW universe.}
 \ee

\subsection{Sunyaev-Zeldovich effect: temperature of scattered CMB photons}

\begin{center}
\begin{figure}[htb!]
\vspace*{-2.5cm}
\includegraphics[width=\columnwidth]{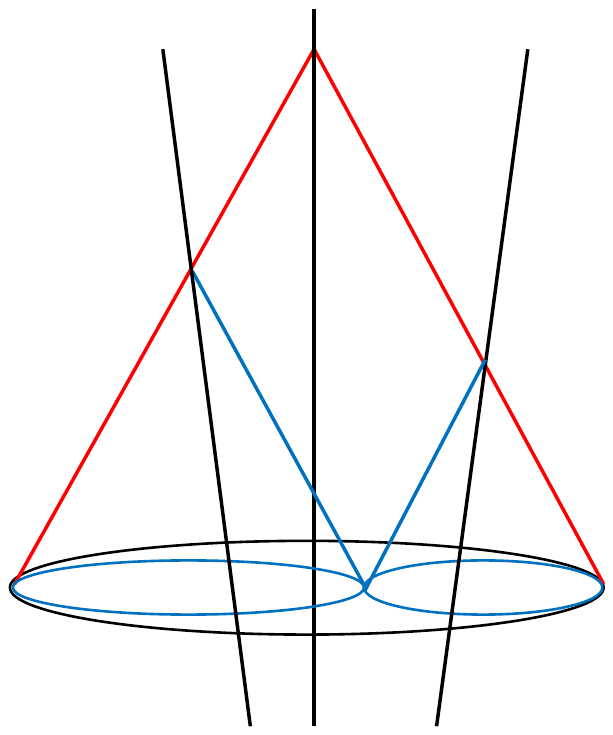}\vspace*{-15.5cm}
\caption{CMB photons from inside our past lightcone are scattered into our line of sight by the ionized gas in galaxy clusters, thus carrying information about the level of isotropy seen by the clusters.
}\label{fig4}
\end{figure}
\end{center}

Galaxy clusters with their hot ionized intra-cluster gas, act via scattering of CMB photons like giant mirrors that carry information about the last scattering surface seen by the cluster (see Fig. \ref{fig4}). In other words, clusters allow us in principle to indirectly probe inside our past lightcone.

CMB photons are scattered into our line of sight, thus inducing spectral distortions in the CMB temperature that we observe \cite{SZ1,SZ2}. The thermal Sunyaev-Zeldovich (SZ) effect from the thermal motion of electrons reflects the monopole seen by the cluster. If the blackbody temperatures of CMB photons arriving at the cluster from points inside our lightcone are significantly different from the blackbody temperature that we directly observe, then there will be a significant distortion. Such a signal would indicate that the cluster sees a significantly anisotropic CMB, hence violating the CP and homogeneity \cite{goodman}. (This test has been applied to a class of LTB models by \cite{CS}.)

The bulk radial motion of the cluster gas induces a kinetic SZ signal that reflects the CMB dipole seen by the cluster. If this is large, then there would be a violation of the CP and homogeneity. This has been applied to classes of LTB models by \cite{gbh2,Zhang:2010fa}, but it is in fact more generally applicable as a test of homogeneity.

In summary,
 \be
\mbox{Large (non-perturbative) thermal or kinetic SZ temperature effect}  ~~\Rightarrow ~~ \mbox{non-FLRW universe.}
 \ee

\subsection{Sunyaev-Zeldovich effect: polarization of scattered CMB photons}

Analogous to the SZ effect on CMB temperature, there is an SZ effect on CMB polarization \cite{Sunyaev:1980nv}. The cluster bulk transverse velocity, and the CMB monopole, quadrupole and octupole, as seen by the cluster, induce modifications in CMB polarization via scattering off the cluster gas. These effects have been computed in FLRW models \cite{Kamionkowski:1997na,Sazonov:1999zp,Challinor:1999yz}. The SZ polarization signals potentially contain more information than the SZ temperature signals, and this has been proposed for the standard FLRW model as a way to map the CMB quadrupole seen at remote locations, thus lessening the cosmic variance \cite{Kamionkowski:1997na,Seto:2000uc,Baumann:2003xb,Bunn:2006mp}.

From a more general viewpoint, the SZ effect on polarization is in principle a powerful probe of the CP and homogeneity. In principle, if CMB polarization data indicate large transverse cluster velocities, or large modifications to the primordial polarization signal, this could signal a violation of the CP and homogeneity:
 \be
\mbox{Large (non-perturbative) SZ polarization effects}  ~~\Rightarrow ~~ \mbox{non-FLRW universe.}
 \ee

\section{CONCLUSIONS}\label{discon}

It is important to re-examine the basic assumptions of the standard concordance model, especially in view of the problems raised by the fine-tuned and unnatural nature of dark energy. Here we have re-examined the central assumption of homogeneity. We argued that homogeneity cannot be established or directly confirmed by observations, given the inherent limitations of lightcone-based data. By contrast, isotropy can be directly probed by observations. We constructed the observational coordinates necessary to answer the question of which observables on the past lightcone are needed to prove isotropy. It turns out that 4 independent matter observables -- angular diameter distances, number counts, lensing distortion and transverse velocities -- are exactly sufficient to determine the spacetime geometry. Isotropy of these 4 observables along one worldline imposes isotropy of the spacetime geometry. Surprisingly, isotropy of the CMB along one worldline does not in itself lead to isotropic geometry.

To establish homogeneity, we are forced to adopt the Copernican Principle (CP). Isotropy of the 4 matter observables for all observers leads to homogeneity -- giving an observational version of the Cosmological Principle. A surprising and more powerful result is that isotropy of only the angular diameter distances for all observers, and only to $O(z^3)$, enforces homogeneity.

The strongest observational basis for homogeneity comes from the CP combined with the high isotropy of the CMB. The main additional assumption needed is that the CMB rest frame is geodesic. We outlined a covariant nonlinear proof and generalization of the original EGS result. And we highlighted the remarkable ETM result, i.e. that the vanishing of the dipole, quadrupole and octupole is sufficient to enforce homogeneity. The realistic case, with almost-isotropy of the CMB, does not lead to almost-homogeneity without additional assumptions on some of the derivatives of the multipoles. It remains an open problem whether these assumptions can be avoided, possibly using further information from almost-isotropy of matter observations.

Although we cannot directly observe homogeneity, we can test homogeneity, using observations that carry information from inside our past lightcone. We described how this can be achieved via consistency relations between distances and expansion rates, using supernova and BAO data, and via Sunyaev-Zeldovich effects from galaxy clusters on the CMB temperature and polarization. Up to now, none of these tests has yet indicated a breakdown of the CP and thus a violation of homogeneity. But the further advance of high-precision data will provide new opportunities to apply and extend these critical tests.

In summary: the standard homogeneous model of cosmology is successful, predictive and up to now robust against all the observational data, and against current tests of homogeneity. The problem of a satisfactory explanation for dark energy may be resolved by advances in particle physics and quantum gravity. It is certainly reasonable to assume that the LCDM model is a good description of the Universe. Nevertheless it is also necessary to continuing probing the foundations of the model -- not only the assumption of homogeneity, but also other critical questions, such as the problem of averaging and the need for a detailed understanding of light propagation in a lumpy Universe.

\[ \]
\[ \]{\bf Acknowledgments:}
 I thank Chris Clarkson,  Catherine Cress, Ruth Durrer, George Ellis, Alan Heavens, Raul Jimenez and Obinna Umeh for discussions. I am supported by a South African SKA Research Chair, by the UK Science \& Technology Facilities Council, and by a Royal Society (UK)/ NRF (South Africa) exchange grant between the Universities of Portsmouth and Western  Cape.

\newpage
\appendix

\section{Nonlinear field and Boltzmann equations}

For convenience we repeat the summary of the 1+3 covariant Lagrangian formulation of the field equations and Boltzmann equation in full nonlinear generality, as given in \cite{ClaMaa10}. For a chosen 4-velocity field $u^\a$, the fundamental tensors are
\begin{equation}\label{hep}
h_{\a\nu}=g_{\a\nu}+u_\a u_\nu,~ ~ \ep_{\a\nu\alpha}=\eta_{\a\nu\alpha\beta}u^\beta,
\end{equation}
where $h_{\a\nu}$ projects into the instantaneous rest space of comoving
observers, and $\ep_{\a\nu\alpha}$ is the projection of the spacetime alternating tensor $\eta_{\a\nu\alpha \beta}=-\sqrt{-g}
\delta^0{}_{[\a}\delta^1{}_\nu\delta^2{}_\alpha\delta^3{}_{\beta]}$.
The projected symmetric tracefree (PSTF) parts of vectors and
rank-2 tensors are
\begin{eqnarray}
V_{\langle \a\rangle}=h_\a{}^\nu V_\nu\,,~ S_{\langle \a\nu \rangle }=
\Big\{h_{(a}{}^\alpha h_{\nu)}{}^\beta-
{{1\over3}}h^{\alpha \beta }h_{\a\nu }\Big\}S_{\alpha \beta }\,. \label{pstf}
\end{eqnarray}
The skew part of a
projected rank-2 tensor is spatially dual to the projected vector,
$S_{\a}={1\over2}\ep_{\a\nu \alpha}S^{[\nu \alpha]}$, and then any projected rank-2
tensor has the decomposition
$S_{\a\nu }={1\over 3}Sh_{\a\nu }+\ep_{\a\nu \alpha}S^{\alpha}+S_{\langle \a\nu \rangle}$, where $S=S_{\alpha\beta} h^{\alpha \beta}$.
The covariant derivative $\nabla_{\a}$ defines 1+3 covariant time and spatial derivatives:
\begin{eqnarray}
\dot{J}^{\a\cdots}{}{}_{\cdots \nu}= u^{\alpha} \nabla_{\alpha}
J^{\a\cdots}{}{}_{\cdots \nu},\, \D_{\alpha} J^{\a\cdots}{}{}_{\cdots \nu} =    h_{\alpha}{}^\beta h^{\a}{}_\kappa\cdots h_{\nu}{}^\tau
\nabla_\beta J^{\kappa\cdots}{}{}_{\cdots \tau}. \label{dd}
\end{eqnarray}
The projected derivative $\D_{\a}$
defines a covariant PSTF divergence, $\D^\a V_\a\,, ~ \D^\nu S_{\a\nu}$,
and a covariant PSTF curl,
\begin{eqnarray}
\mbox{curl}\, V_{\a}=\ep_{\a\nu c}\D^{\nu}V^{\alpha}\,,~ \mbox{curl}\,
S_{\a\nu }=\ep_{\alpha \beta (\a}\D^{\alpha}S_{\nu)}{}^\beta\,. \label{curl}
\end{eqnarray}

The relative motion of comoving observers
is encoded in the PSTF kinematical quantities: the volume expansion rate,  4-acceleration, vorticity
and shear, given respectively by
\begin{eqnarray}
\Theta=\D^{\a}u_{\a},~ A_{\a}=\dot{u}_{\a},~ {\omega}_\a=\mbox{curl}\, u_\a ,~ \sigma_{\a\nu }=\D_{\langle \a}u_{\nu\rangle }~ \Leftrightarrow~ \nabla_{\nu}u_{\a}={{1\over3}}\Theta h_{\a\nu }+\ep_{\a\nu \alpha}\omega^{\alpha}
+\sigma_{\a\nu }-A_{\a}u_{\nu}\,. \label{kin}
\end{eqnarray}
Key nonlinear identities are
\begin{eqnarray}
\mbox{curl}\,\D_{\a}\psi & := & \ep_{\a\nu \alpha}\D^{\nu}\D^{\alpha}\psi=
-2\dot{\psi}\omega_{\a} \,, \label{ri1}\\
h_\a^{~\nu}(\D_{\nu}\psi\dot)-\D_{\a}\dot\psi &=& \dot\psi A_\a-\left(\frac{1}{3}\Theta h_{\a\nu}+\sigma_{\a\nu}+\varepsilon_{\a\nu \kappa}\omega^\kappa\right)\D^\nu\psi\, .\label{timespace}
\end{eqnarray}

The PSTF dynamical quantities describe the
sources of the gravitational field:
the (total) energy density $\rho=T_{\a\nu }u^{\a}u^{\nu}$ and isotropic pressure
$p={1\over3}h_{\a\nu }T^{\a\nu }$ (including $\Lambda$), momentum density $q_{\a}=-T_{\langle \a\rangle \nu}u^{\nu}$,
and anisotropic stress $\pi_{\a\nu }=T_{\langle \a\nu \rangle}$, where $T_{\a\nu }$ is
the total energy-momentum tensor. The Weyl tensor
splits into PSTF gravito-electric and gravito-magnetic fields
\begin{eqnarray}
E_{\a\nu }=C_{\a \alpha \nu \beta}u^{\alpha}u^\beta\,,~~
H_{\a\nu }={{1\over2}}\ep_{\a\alpha \beta }C^{\alpha \beta }{}{}_{\nu \kappa}u^\kappa  \,.\label{gem}
\end{eqnarray}
The Ricci and Bianchi identities,
\begin{eqnarray}
\nabla_{[\a} \nabla_{\nu]} u_\alpha= R_{\a\nu \alpha  \beta}u^{\beta}, ~~~ \nabla^\beta
C_{\a\nu \alpha \beta } =- \nabla_{[\a}\Big\{ R_{\nu]\alpha} - {1\over6}Rg_{\nu]\alpha} \Big\}, \label{rbi}
\end{eqnarray}
produce the
fundamental evolution and constraint equations governing the
covariant quantities. Einstein's equations are
incorporated via the algebraic replacement $
R^{\a\nu }=T^{\a\nu }-{1\over2}T_{\alpha}{}^{\alpha}g^{\a\nu }
$. The resulting equations, in fully nonlinear form and for a general
source of the gravitational field, are:

\noindent{\em Evolution:}
\begin{eqnarray}
&& \dot{\rho} +(\rho+p)\Theta+\D^\a q_\a = -2A^{\a} q_{\a}
 -\sigma^{\a\nu }\pi_{\a\nu }\,, 
 \label{e1}\\
&& \dot{\Theta} +{{1\over3}}\Theta^2 +{{1\over2}}(\rho+3p)-\D^\a
A_\a
= -\sigma_{\a\nu }\sigma^{\a\nu }
+2\omega_{\a}\omega^{\a}+A_{\a}A^{\a} \,,
\label{e2}\\
&& \dot{q}_{\langle \a\rangle }
+{{4\over3}}\Theta q_{\a}+(\rho+p)A_{\a} +\D_{\a} p +\D^\nu\pi_{\a\nu}
= -\sigma_{\a\nu }q^{\nu}
+\ep_{\a\nu\alpha}\omega^\nu q^{\alpha} -A^{\nu}\pi_{\a\nu } \,,
\label{e3} \\
&& \dot{\omega}_{\langle \a\rangle } +{{2\over3}}\Theta\omega_{\a}
+{{1\over2}}\mbox{curl}\, A_{\a} = \sigma_{\a\nu }\omega^{\nu} \,,\label{e4}\\
&& \dot{\sigma}_{\langle \a\nu \rangle } +{{2\over3}}\Theta\sigma_{\a\nu }
+E_{\a\nu }-{{1\over2}}\pi_{\a\nu } -\D_{\langle \a}A_{\nu\rangle } 
= -\sigma_{\alpha\langle \a}\sigma_{\nu\rangle }{}^{\alpha} - \omega_{\langle \a}\omega_{\nu\rangle }
+A_{\langle \a}A_{\nu\rangle }\,,
\label{e5}\\
&& \dot{E}_{\langle \a\nu \rangle } +\Theta E_{\a\nu }
-\mbox{curl}\, H_{\a\nu } +{{1\over2}}(\rho+p)\sigma_{\a\nu } 
+{{1\over2}}
\dot{\pi}_{\langle \a\nu \rangle } +{{1\over6}}
\Theta\pi_{\a\nu } +{{1\over2}}\D_{\langle \a}q_{\nu\rangle } \nonumber\\
&&~~~~~~~ =-A_{\langle \a}q_{\nu\rangle } +2A^{\alpha}\ep_{\alpha \beta (\a}H_{\nu)}{}^\beta
+3\sigma_{\alpha\langle \a}E_{\nu\rangle }{}^{\alpha}
-\omega^{\alpha} \ep_{\alpha \beta (\a}E_{\nu)}{}^\beta -{{1\over2}}\sigma^{\alpha}{}_{\langle
\a}\pi_{\nu\rangle \alpha} -{{1\over2}}\omega^{\alpha}\ep_{\alpha \beta (\a}\pi_{\nu)}{}^\beta \,,
\label{e6}\\
&& \dot{H}_{\langle \a\nu \rangle } +\Theta H_{\a\nu } +\mbox{curl}\, E_{\a\nu }
-{{1\over2}}\mbox{curl}\,\pi_{\a\nu }
= 3\sigma_{\alpha\langle \a}H_{\nu\rangle }{}^{\alpha}
-\omega^{\alpha} \ep_{\alpha \beta (\a}H_{\nu)}{}^\beta
\nonumber\\
&&~~~~~~~~~
-2A^{\alpha}\ep_{\alpha \beta (\a}E_{\nu)}{}^\alpha -{{3\over2}}\omega_{\langle \a}q_{\nu\rangle
}+ {{1\over2}}\sigma^{\alpha}{}_{(\a}\ep_{\nu)\alpha \beta }q^\beta \,. \label{e7}
\end{eqnarray}

\noindent{\em Constraint:}
\begin{eqnarray}
&& \D^\a\omega_\a = A^{\a}\omega_{\a} \,, \label{c1}\\
&& \D^\nu\sigma_{\a\nu}-\mbox{curl}\,\omega_{\a} -{{2\over3}}\D_{\a}\Theta +q_{\a} =-
2\ep_{\a\nu\alpha}\omega^\nu A^{\alpha}  \,,\label{c2}\\
&&  \mbox{curl}\,\sigma_{\a\nu }+\D_{\langle \a}\omega_{\nu\rangle }
 -H_{\a\nu }= -2A_{\langle \a}
\omega_{\nu\rangle } \,,\label{c3}\\
&& \D^\nu E_{\a\nu}
+{{1\over2}}\D^\nu\pi_{\a\nu}
 -{{1\over3}}\D_{\a}\rho
+{{1\over3}}\Theta q_{\a}
= \ep_{\a\nu\alpha}\sigma^\nu{}_\beta H^{\alpha \beta} -3H_{\a\nu}
\omega^{\nu} +{{1\over2}}\sigma_{\a\nu }q^{\nu}-{{3\over2}}
\ep_{\a\nu\alpha}\omega^\nu q^{\alpha}  \,,\label{c4}\\
&& \D^\nu H_{\a\nu}
+{{1\over2}}\mbox{curl}\, q_{\a}
 -(\rho+p)\omega_{\a}
=-\ep_{\a\nu\alpha}\sigma^\nu{}_\beta E^{\alpha \beta}-{{1\over2}}\ep_{\a\nu\alpha}\sigma^\nu{}_\beta \pi^{\alpha \beta}   +3E_{\a\nu }\omega^{\nu}
-{{1\over2}}\pi_{\a\nu } \omega^{\nu}  \,.\label{c5}
\end{eqnarray}

The energy and momentum conservation equations are (\ref{e1}) and (\ref{e3}). The total dynamical quantities have
contributions from all dynamically significant particle species:
\begin{eqnarray}
T^{\a\nu } = \sum_I T_{I}^{\a\nu } = \rho
u^{\a}u^{\nu}+ph^{\a\nu }+2q^{(\a}u^{\nu)}
+\pi^{\a\nu } \,, ~~ 
T_{I}^{\a\nu }=
\rho^*_{I}u_{I}^{\a}u_{I}^{\a}+p^*_{I}h_{I}^{\a\nu }
+2q_{I}^{*(\a}u_{I}^{\nu)}+\pi_{I}^{*\a\nu }\,, \label{t2}
\end{eqnarray}
where $I=\mbox{r,n,b,c},\Lambda$ labels the species. The asterisk on the
dynamical quantities $\rho^*_{I},\cdots$ denotes that these quantities are measured, not in the $u^\a$-frame, but in the $I$-frame, whose 4-velocity is given by
\begin{eqnarray}
u_{I}^{\a}=\gamma_{I}\left(u^{\a}+v_{I}^{\a}\right)\,, ~v_{I}^{\a}u_{\a}=0\,, ~\gamma_I=\left( 1-v_I^2 \right)^{-1/2}.\label{t3}
\end{eqnarray}
The fully nonlinear equations for the $I$ dynamical quantities as measured in the fundamental $u^{\a}$-frame are:
\begin{eqnarray}
\rho_{I} &=& \rho^*_{I} +
\Big\{\gamma_{I}^2v_{I}^2\left(\rho^*_{I}+p^*_{I}\right)
+2\gamma_{I}q_{I}^{*\a}
v_{I\a}+\pi_{I}^{*\a\nu }v_{I\a}v_{I\nu}\Big\} \,,\label{t4}\\
p_{I} &=&  p^*_{I} +{{1\over3}}
\Big\{\gamma_{I}^2v_{I}^2\left(\rho^*_{I}
+p^*_{I}\right)+2\gamma_{I}q_{I}^{*\a}
v_{I\a}+\pi_{I}^{*\a\nu }v_{I\a}v_{I\nu}\Big\}\,, \label{t5}\\
q_{I}^{\a} &=& q_{I}^{*\a}+(\rho^*_{I}+p^*_{I})v_{I}^{\a}
+\Big\{ (\gamma_{I}-1)q_{I}^{*\a}
-\gamma_{I}q_{I}^{*\nu}v_{I\nu}u^{\a}+ \gamma_I q^{*\nu}_I v_{I\nu} v_I^\mu\nonumber\\
&&{} +\gamma_{I}^2v_{I}^2
\left(\rho^*_{I}+p^*_{I}\right)v_{I}^{\a}
+\pi_{I}^{*\a\nu }v_{I\nu}-\pi_{I}^{*\nu\alpha} v_{I\nu}v_{I\alpha}u^{\a}
\Big\} \,,
\label{t6}\\
\pi_{I}^{\a\nu } &=& \pi_{I}^{*\a\nu } +
\Big\{-2u^{(\a}\pi_{I}^{*\nu)\alpha}v_{I\alpha}+\pi_{I}^{*\nu\alpha} v_{I\nu}
v_{I\alpha}u^{\a}u^{\nu} +2\gamma_{I}v_{I}^{\langle \a}q_{I} ^{*\nu\rangle} \nonumber\\
&&{}- 2 \gamma_I q^{*\alpha}_I v_{I\alpha} u^{(\mu} v_I^{\nu)} -{1\over 3}\pi_{I}^{*\alpha \beta }v_{I\alpha}v_{I\beta}h^{\a\nu } +
\gamma_{I}^2\left(\rho^*_{I}+p^*_{I}\right) v_{I}^{\langle
\a}v_{I}^{\nu\rangle}
\Big\}\,. \label{t7}
\end{eqnarray}
The terms in braces are the nonlinear corrections that vanish in the standard perturbed FLRW case.
The total dynamical quantities in (\ref{e1})--(\ref{c5}), are given by
\begin{eqnarray}
\rho=\sum_I\rho_{I}\,,~p=\sum_I p_{I}\,,~ q^{\a}=\sum_I
q_{I}^{\a}\,,~\pi^{\a\nu }=\sum_I\pi_{I}^{\a\nu }\,.
\end{eqnarray}
Assuming that the species are non-interacting, they each separately obey the energy and momentum conservation equations (\ref{e1}) and (\ref{e3}):
\begin{eqnarray}
&& \dot{\rho}_I +(\rho_I+p_I)\Theta+\D_\a q_I^\a = -2A_{\a} q_I^{\a}
 -\sigma_{\a\nu }\pi_I^{\a\nu }\,, 
 \label{e1i}\\
&& \dot{q}_I^{\langle \a\rangle }
+{{4\over3}}\Theta q_I^{\a}+(\rho_I+p_I)A^{\a} +\D^{\a} p_I +\D_\nu\pi_I^{\a\nu}
= -\sigma^\a{}_\nu q_I^{\nu}
+\ep^\a{}_{\nu\alpha}\omega^\nu q_I^{\alpha} -A_{\nu}\pi_I^{\a\nu } \,,
\label{e3i}
\end{eqnarray}
where the $I$-quantities are given by (\ref{t4})--(\ref{t7}).

The covariant kinetic theory description starts by splitting the photon 4-momentum as
\begin{equation}\label{ph4m}
p^{\a}=E(u^{\a}+e^{\a})\,,~~e^{\a} e_{\a}=1\,,~e^{\a} u_{\a}=0\,. 
\end{equation}
Here $E=-u_{\a}p^{\a}$ is the energy and $e^{\a}=p^{\langle \a\rangle}/E$ is the
direction, as measured by a comoving fundamental observer. Then
the photon distribution function is decomposed into covariant
harmonics via the expansion
\begin{eqnarray}
f(x,p)=f(x,E,e) &=& F+F_{\a}e^{\a}+F_{\a\nu }e^{\a}e^{\nu}+\cdots 
=\sum_{\ell\geq0}
F_{M_\ell}(x,E) e^{\langle M_\ell\rangle}, \label{r3}
\end{eqnarray}
where
${M_\ell}:= {\a_1}{\a_2}\cdots {\a_\ell}$ and $e^{M_\ell} := e^{\a_1}\cdots e^{\a_\ell}$. The PSTF multipoles $F_{M_\ell}$ are a covariant alternative to the usual expansion in spherical harmonics. The energy-momentum
tensor is $
T_{\rm r}^{\a\nu }(x)=\int p^{\a}p^{\nu}f(x,p)\mathrm{d}^3p$, so the first 3 multipoles define the radiation dynamical quantities (in the $u^{\a}$-frame):
\begin{eqnarray}
\rho_{\rm r} = 4\pi\int_0^\infty E^3F\,\mathrm{d} E\,, ~q_{\rm r}^{\a} =
{4\pi\over 3}\int_0^\infty E^3F^{\a}\,\mathrm{d} E \,,~
\pi_{\rm r}^{\a\nu }=
{8\pi\over 15}\int_0^\infty E^3F^{\a\nu }\,\mathrm{d} E\,. \label{em3} \end{eqnarray}
Thus $\Pi= (1/4\pi)\rho_{\rm r}, \Pi^{\a}=(3/ 4\pi)q_{\rm r}^{\a}, \Pi^{\a\nu }=(15/ 8\pi)\pi_{\rm r}^{\a\nu }$, where the brightness multipoles are
\begin{eqnarray}
\Pi_{\a_1\cdots \a_\ell} = \int E^3
F_{\a_1\cdots \a_\ell}\mathrm{d} E\,. \label{r10}
\end{eqnarray}

The collisionless Boltzmann (or Liouville) equation is
\begin{equation}
{\mathrm{d} f\over \mathrm{d} v}:= p^{\a}{\p f\over \p x^{\a}}-\Gamma^{\a}{}_{\alpha\beta}
p^{\alpha}p^{\beta}{\p f\over \p p^{\a}}=0 \,, \label{boltz}
\end{equation}
where $p^{\a}=\mathrm{d} x^{\a}/\mathrm{d} v$, and we neglect polarization.
The covariant multipoles of $\mathrm{d} f/\mathrm{d} v$ are given by
\begin{eqnarray}
&&{1\over E}\left({\mathrm{d} f \over \mathrm{d} v}\right)_{M_\ell} =\dot{F}_{\langle M_\ell \rangle}-{{1\over3}}\Theta
EF'_{M_\ell} +\D_{\langle \a_\ell} F_{M_{\ell-1}\rangle}
+{(\ell+1)\over(2\ell+3)} \D^{\a}F_{\a M_\ell} ~~~~
\nonumber\\
&&{}
-{(\ell+1)\over(2\ell+3)}E^{-(\ell+1)}\left[E^{\ell+2}F_{\a M_\ell}
\right]'A^{\a}-E^\ell\left[E^{1-\ell}F_{\langle M_{\ell-1}}\right]'
A_{\a_\ell\rangle}
\nonumber\\
&&{} -\ell\omega^{\nu}\ep_{\nu\alpha( \a_\ell} F_{M_{\ell-1})}{}^{\alpha}
-{(\ell+1)(\ell+2)\over(2\ell+3)(2\ell+5)}E^{-(\ell+2)}
\left[E^{\ell+3}F_{\a\nu M_\ell}\right]'\sigma^{\a\nu }
\nonumber\\
&&{} -{2\ell\over (2\ell+3)}E^{-1/2}\left[E^{3/2}F_{\nu\langle
M_{\ell-1}}
\right]'\sigma_{\a_\ell\rangle}{}^{\nu}
 -E^{\ell-1}\left[E^{2-\ell} F_{\langle
M_{\ell-2}}\right]'\sigma_{\a_{\ell-1}\a_\ell\rangle}\,,
\label{r25}\end{eqnarray}
where a prime denotes $\p/\p E$. This is a fully nonlinear expression.
Multiplying by $E^3$ and integrating over all
energies leads to the brightness multipole evolution equations:
\begin{eqnarray}
0 &=& \dot{\Pi}_{\langle M_\ell\rangle}+{{4\over3}}\Theta
\Pi_{M_\ell}+ \D_{\langle \a_\ell}\Pi_{M_{\ell-1}\rangle}
+{(\ell+1)\over(2\ell+3)}\D^{\nu} \Pi_{\nu M_\ell}
\nonumber\\
&&{} -{(\ell+1)(\ell-2)\over(2\ell+3)} A^{\nu} \Pi_{\nu M_\ell} +(\ell+3)
A_{\langle \a_\ell} \Pi_{M_{\ell-1}\rangle}\nonumber\\
&&{} -\ell\omega^{\nu}\ep_{\nu\alpha( \a_\ell}
\Pi_{M_{\ell-1})}{}^{\alpha}
 -{(\ell-1)(\ell+1)(\ell+2)\over(2\ell+3)(2\ell+5)}
\sigma^{\nu\alpha}\Pi_{\nu\alpha M_\ell} \nonumber\\
&&{} +{5\ell\over(2\ell+3)} \sigma^{\nu}{}_{\langle
\a_\ell} \Pi_{M_{\ell-1}\rangle \nu} -(\ell+2) \sigma_{\langle
\a_{\ell}\a_{\ell-1}} \Pi_{M_{\ell-2}\rangle}\,.
\label{r26}\end{eqnarray}
The monopole equation is just the energy conservation equation, i.e., (\ref{e1i}) with $I=r$, the dipole is the momentum conservation equation (\ref{e3i}), with $I=r$, and the quadrupole is (\ref{nl8}).

\end{document}